\documentstyle[12pt]{article}
\catcode`@=11
\def\seceqa{\@addtoreset{equation}{section}
           \def\theequation{1.\arabic{equation}}}
\def\seceqb{\@addtoreset{equation}{section}
           \def\theequation{2.\arabic{equation}}}
\def\seceqc{\@addtoreset{equation}{section}
           \def\theequation{3.\arabic{equation}}}
\def\seceqd{\@addtoreset{equation}{section}
           \def\theequation{4.\arabic{equation}}}
\def\seceqe{\@addtoreset{equation}{section}
           \def\theequation{5.\arabic{equation}}}
\def\seceqf{\@addtoreset{equation}{section}
           \def\theequation{6.\arabic{equation}}}
\def\seceqaa{\@addtoreset{equation}{section}
           \def\theequation{A\arabic{equation}}}
\def\seceqab{\@addtoreset{equation}{section}
           \def\theequation{B\arabic{equation}}}
\def\seceqac{\@addtoreset{equation}{section}
           \def\theequation{C\arabic{equation}}}
\def\seceqad{\@addtoreset{equation}{section}
           \def\theequation{D\arabic{equation}}}
\catcode`@=12

\def\lsim{<\kern-2.5ex\lower0.85ex\hbox{$\sim$}\ }
\def\rsim{>\kern-2.5ex\lower0.85ex\hbox{$\sim$}\ }

\begin{document}

\vskip 1.5 true in
\title{Inelastic nucleon contributions in $(e,e^\prime)$ nuclear response
functions}
\author{T.C.~Ferr\'ee and D.S.~Koltun \\
{\it Department of Physics and Astronomy}\\
{\it University of Rochester}\\
{\it Rochester, New York 14627-0171}}
\maketitle
\vskip 0.5 true in

\begin{abstract}
We estimate the contribution of inelastic nucleon excitations to the
$(e,e^\prime)$ inclusive cross section in the CEBAF kinematic range.
Calculations are based upon parameterizations of the nucleon structure
functions measured at SLAC.  Nuclear binding effects are included in
a vector-scalar field theory, and are assumed have a minimal effect
on the nucleon excitation spectrum.  We find that for $q\lsim 1$ GeV
the elastic and inelastic nucleon contributions to the nuclear response
functions are comparable, and can be separated, but with roughly a
factor of two uncertainty in the latter from the extrapolation from
data.  In contrast, for $q\rsim 2$ GeV this uncertainty is greatly
reduced but the elastic nucleon contribution is heavily dominated by
the inelastic nucleon background.
\hfill\break
\hfill\break
\end{abstract}
\vskip 0.5 true in
\centerline{Submitted to {\em Physical Review C}}
\vfil
\eject

\section{Introduction}
\setcounter{equation}{0}
\seceqa

Inelastic scattering of electrons from nuclear targets has long been a
tool for the study of nuclear structure.  For example, the Coulomb
response has been experimentally determined with a view to testing the
Coulumb sum rule against models of nuclear structure.  Most data and
analyses have been obtained for three-momentum transfers $q<550{\rm\
MeV/c}$,\cite{xa}--\cite{xg} although some recent work has been done
for $q\sim1{\rm\ GeV/c}$.\cite{mez}  At the lower $q$ and energy transfers
$\omega$ ($<q$) the nuclear response involves almost entirely nucleons
without internal excitation:  the $\Delta$-resonance is excited at
higher $\omega$, but contributes mainly to the transverse response,
and therefore has less effect on the Coulomb response.

Interest in extending the study of the nuclear $(e,e^\prime)$ response
to higher $q$ has developed in recent years, especially with the advent
of the Continuous Electron Beam Accelerator Facility (CEBAF), with beam
energies of $\omega \sim 4{\rm\ GeV}$ or more.  This has led to the study
of relativistic effects in the nuclear response which become important for
$q\rsim1{\rm\ GeV/c}$.\cite{wal}--\cite{kf}  But for such momentum transfers
the probability of exciting internal states of the {\em nucleons} becomes
increasingly important.  It is the purpose of this paper to make an
estimate of the magnitude of the contribution of internal excited states
of the nucleon to the nuclear response based on measured values of the
inelastic nucleon structure functions and a Fermi gas model of the nuclear
target.  The contribution of inelastically excited nucleons may be
considered the background to the nuclear response function with non-excited
(elastic) nucleons; it is the latter which is usually compared to models
of nuclear structure.  Our estimates are for $1\le q\le4{\rm\ GeV/c}$,
with the appropriate ranges of $\omega$.

Electroexcitation of nucleons has been studied at the Stanford Linear
Accelerator (SLAC) and elsewhere in this momentum range, but more recently
at much higher momenta.  We make use of systematic fits\cite{bodek,whitlow}
to SLAC data, which include some of our range of interest, or lie close
enough for extrapolation.  We ignore the details of the low energy
resonances \cite{stein} in the relevant range, which are only
approximately included in one of the parameterizations\cite{bodek}
and not at all in the second\cite{whitlow}.  We concentrate on the
magnitude and shape of the background from smoothly varying fits to the
nucleon structure functions.  Resonances should of course be added, but
this will require more complete data separating the longitudinal and
transverse contributions.

The paper begins with a review of the basic formalism for $(e,e^\prime)$
on complex targets, which defines the structure (or response) functions
for nucleon or nuclear targets.  In Section 3 we formulate a simple
impulse model for the response of a nuclear target based on a Fermi gas
of bound nucleons.  This follows earlier treatments of Fermi smearing,
e.g., by Bodek and Ritchie,\cite{br} but with some detailed differences
in the prescription for going off shell.  The treatment of binding effects
continues in Section 4.  The parametric fits to the nucleon structure
functions appear in Section 5.  Finally, we present our estimates for
the contribution of nucleon inelastic (internally excited) response
functions in Section 6, and discuss these results in Section 7.

\section{Basic formalism}
\setcounter{equation}{0}
\seceqb

In this section we summarize the basic formalism for $(e,e^\prime)$
scattering from a nuclear target, which may be either a single nucleon
or an $A$-body nucleus.  The following sections give extensions for bound
nucleons.  In the Born (one-photon exchange) approximation one can write
the differential cross section

	\begin{equation}{d^2\sigma\over d\Omega^\prime dE^\prime}=
	{\alpha^2\over q^4}{|{\bf k}^\prime|\over|{\bf k}|}
	L_e^{\mu\nu}W_{\mu\nu}\ ,\label{dcs1}\end{equation}

\noindent where $L_e^{\mu\nu}$ is a lepton tensor describing incoming and
outgoing plane-wave electron states, $k$ and $k^\prime$ represent the
initial and final electron four-momenta, and $q\equiv k-k^\prime$ is
the four-momentum transferred to the target via virtual photon exchange.

Let $p$ and $p^\prime$ represent the initial and final four-momenta of
the target.  In principle the target (response) tensor $W_{\mu\nu}$ is
a function of all three variables $p$, $p^\prime$ and $q$, however,
conservation of four-momentum can be used to eliminate any one of these;
it is conventional to eliminate $p^\prime$.  The most general form of
$W_{\mu\nu}$ satisfying Lorentz invariance, gauge invariance and parity
can then be written\cite{drell}

        \begin{equation}W_{\mu\nu}(p,q)=W_1\Biggl[-g_{\mu\nu}
        +{q_\mu q_\nu\over q^2}\Biggr]+W_2
        \Biggl[{p_\mu\over M}-{p\cdot q\over M}{q_\mu\over q^2}\Biggr]
        \Biggl[{p_\nu\over M}-{p\cdot q\over M}{q_\nu\over
        q^2}\Biggr]\ ,\label{wuv}\end{equation}

\noindent where the tensor behavior of $W_{\mu\nu}$ under Lorentz
transformations is described entirely by the quantities in square brackets.
The scalar structure functions $W_1$ and $W_2$ depend only on scalar
combinations of $p$ and $q$ (and the target mass $M$), from which one can
form three scalar combinations: $p^2$, $q^2$ and $p\cdot q$, or equivalently
$Q^2\equiv-q^2$, $\nu\equiv p\cdot q/M$ and $W^2\equiv(p+q)^2$.  For an
on-shell target $p^2=M^2$ so only two independent variables remain.  A
conventional choice for nucleons, which we adopt, is $Q^2$ and $\nu$.

For a nuclear target at rest in the laboratory frame, inserting
(\ref{wuv}) into (\ref{dcs1}) leads to

        \begin{equation}{d^2\sigma\over d\Omega^\prime dE^\prime}=
        {d\sigma_{\!_M}\over d\Omega^\prime}\
        \Biggl[{Q^2\over{\bf q}^2}W_L^A(\omega,{\bf q})
        +\biggl({1\over2}{Q^2\over{\bf q}^2}+{\rm tan}^2
        {\theta\over2}\biggr)W_T^A(\omega,{\bf q})\Biggr]\ ,
        \label{dcs2}\end{equation}

\noindent where the superscript $A$ refers to the $A$-body nucleus.  The
longitudinal ($W_L^A$) and transverse ($W_T^A$) structure (or response)
functions are defined

        \begin{equation}W_L^A(\omega,{\bf q})\equiv{{\bf q}^2\over Q^2}
        W_2^A(\omega,{\bf q})-W_1^A(\omega,{\bf q})\ ,\label{wldef}
        \end{equation}

        \begin{equation}W_T^A(\omega,{\bf q})\equiv2\times
        W_1^A(\omega,{\bf q})\ .\label{wtdef}\end{equation}

\noindent Equations (\ref{dcs2})--(\ref{wtdef}) have been written as
functions of the lab variables $\omega$ (energy transfer) and ${\bf q}$
(three-momentum transfer) with $q^\mu=(\omega,{\bf q})$, as is standard
for analyzing the nuclear response.  In these expressions the target mass
$M$ which appears in (\ref{wuv}) has dropped out since $p_0=M$ for an
on-shell target at rest.  Note also that in this case $\nu=\omega$.

Since real photons are purely transverse, $W_L^A=0$ at $Q^2=0$, i.e.,
at $\omega=|{\bf q}|$.  In order to analyze nuclear $(e,e^\prime)$ data
for the purpose of extracting two-body correlation functions\cite{fk,kf}
the function $W_L^A$ must be evaluated over the entire range
$0\le\omega\le|{\bf q}|$.  Its behavior near $Q^2=0$ is therefore an
important qualitative feature which must be maintained.

\section{Fermi smearing}
\setcounter{equation}{0}
\seceqc

We are primarily interested in estimating the contribution of internal
excited states of the nucleon to the inelastic nuclear response, compared
to the nuclear response in which nucleons are not excited internally.
For this comparison we need a model in which we can express the $A$-body
target tensor $W_{\mu\nu}^A(\omega,{\bf q})$ in terms of
$W_{\mu\nu}^\sigma(p,q)$, the corresponding $1$-body tensor for
constituent nucleons in the target of isospin projection $\sigma$.
We also need quantitative information on the nucleon structure function,
for both elastic and inelastic {\em nucleon} kinematics.

Any realistic model which describes an $A$-body nucleus in terms of its
constituent nucleons must include at least Fermi motion and nuclear
binding effects.  The simplest model is the plane-wave impulse
approximation (PWIA) in which final state interactions are ignored.
This should be adequate to provide a reasonable estimate of the inelastic
nucleon background.  Within the PWIA, West\cite{west} has shown that
the $A$-body response tensor in the laboratory is given by
\footnote{For elastic nucleons Pauli blocking in the final state
should be included, and leads to an additional factor
$\bigl[1-n_\sigma({\bf p+q})\bigr]$.  See (\ref{wlqe}) and
(\ref{wtqe}).}

        \begin{equation}W^A_{\mu\nu}(\omega,{\bf q})=2\ \sum_\sigma
        \int{d^3p\over(2\pi)^3}{n_\sigma({\bf p})\over E_{\bf p}/M}
        W_{\mu\nu}^\sigma(p,q)\ ,\label{smear}\end{equation}

\noindent where $E_{\bf p}\equiv\sqrt{{\bf p}^2+M^2}$ and $M$ is the
nucleon mass.  The momentum distribution $n_\sigma({\bf p})$ for
nucleon species $\sigma$ is normalized to $N_\sigma/2$, where $N_p
=Z$ and $N_n=N$, and the overall factor of $2$ reflects a sum over
both spin states.  The factor $E_{\bf p}/M$ is required to preserve
the phase space volume of $d^3p$ under Lorentz
transformations\cite{awest}.

For a free Fermi gas we would put (\ref{wuv}) on the right-hand side of
(\ref{smear}), however, expression (\ref{wuv}) was derived for on-shell
targets and is not strictly valid for bound nucleons.  A complete
specification of $W^\sigma_{\mu\nu}$ for the general off-shell case is
outside the scope of this paper because it requires knowledge of nucleon
dynamics in the medium.  We will simply assume that (\ref{wuv}) is a
valid starting point for an off-shell extension.  Binding effects
then enter $W_{\mu\nu}^\sigma$ in two ways: 1) through the initial-state
nucleon energy $p_0$ which appears explicitly in the square brackets in
(\ref{wuv}), and 2) implicitly through the scalar functions $W_1^\sigma$
and $W_2^\sigma$.  We have already stated that on shell $W_1^\sigma$ and
$W_2^\sigma$ are functions only two variables, which we take to be $Q^2$
and $\nu$, and that for a bound nucleon $W_1^\sigma$ and $W_2^\sigma$ can
in principle depend on three variables, e.g., $Q^2$, $\nu$ and $W^2$.
Since experimental data for $W_{\mu\nu}^\sigma$ is available only on shell,
\footnote{Technically speaking, the proton tensor $W_{\mu\nu}^p$ is
measured on shell by scattering from free protons, while the neutron
tensor $W_{\mu\nu}^n$ is determined by scattering from deuterons and
extracting neutron contributions using theoretical arguments.}  a
minimal procedure is to assume that off shell $W_1^\sigma$ and
$W_2^\sigma$ are also functions of only two variables.  In principle
there exists considerable freedom to choose which two, however, in what
follows we will show that $Q^2$ and $\nu$ is the natural choice in
order to satisfy the requirement that $W_L^A=0$ at $Q^2=0$.

To procede we insert (\ref{wuv}) into each side of (\ref{smear}).
With the $z$-axis chosen along ${\bf q}$ and assuming spherical
symmetry for $n({\bf p})$, equating tensor components leads to

        \begin{equation}W^A_1(\omega,{\bf q})=2\ \sum_\sigma\int
        {d^3p\over(2\pi)^3}{n_\sigma({\bf p})\over E_{\bf p}/M}
        \Biggl[W_1^\sigma(Q^2,\nu)
        +\biggl({p_x\over M}\biggr)^2
        W_2^\sigma(Q^2,\nu)\Biggr]\ ,\label{w1a}
        \end{equation}

        \begin{equation}W^A_2(\omega,{\bf q})=2\ \sum_\sigma\int
        {d^3p\over(2\pi)^3}{n_\sigma({\bf p})\over E_{\bf p}/M}
        \Biggl[\biggl(1+{p_z\over M}{Q^2\over|{\bf q}|\nu}\biggr)^2
        \biggl({\nu\over\omega}\biggr)^2+\biggl({p_x\over M}\biggr)^2
        {Q^2\over{\bf q}^2}\Biggr]W_2^\sigma
        (Q^2,\nu)\ .\label{w2a}\end{equation}

\noindent Inserting (\ref{w1a}) and (\ref{w2a}) into (\ref{wldef})
and (\ref{wtdef}) leads to

        \begin{equation}W_L^A(\omega,{\bf q})=2\ \sum_\sigma\int
        {d^3p\over(2\pi)^3}{n_\sigma({\bf p})\over E_{\bf p}/M}
        \Biggl[{{\bf q}^2\over Q^2}
        \biggl(1+{p_z\over M}{Q^2\over|{\bf q}|\nu}\biggr)^2
        \biggl({\nu\over\omega}\biggr)^2\ W_2^\sigma(Q^2,\nu)
        -W_1^\sigma(Q^2,\nu)\Biggr]\ ,
        \label{wla}\end{equation}

        \begin{equation}W^A_T(\omega,{\bf q})=2\times 2\ \sum_\sigma
        \int{d^3p\over(2\pi)^3}{n_\sigma({\bf p})\over E_{\bf p}/M}
        \Biggl[W_1^\sigma(Q^2,\nu)+\biggl({p_x\over M}\biggr)^2
        W_2^\sigma(Q^2,\nu)\Biggr]\ .
        \label{wta}\end{equation}

\noindent From ({\ref{wla}) it is easily verified that the condition
$W_L^A=0$ at $Q^2=0$ will be satisfied if

        \begin{equation}\lim_{Q^2\rightarrow0}
	\Biggl[{\nu^2\over Q^2}W_2^\sigma(Q^2,\nu)
	-W_1^\sigma(Q^2,\nu)\Biggr]=0\ .
	\label{crucial}\end{equation}

\noindent  To understand the implications of (\ref{crucial}) we first
observe that the factor $\nu^2/Q^2$ originates from the $p\cdot q$ terms
in (\ref{wuv}).  Although the choice of variables on which $W_1^\sigma$
and $W_2^\sigma$ depend off shell is in principle arbitrary, this work
is based on the assumptions that 1) $W_1^\sigma$ and $W_2^\sigma$ depend
on {\em only} two variables, and 2) $W_1^\sigma$ and $W_2^\sigma$ depend
on the {\em same~} two variables.  Consequently (\ref{crucial}) will be
satisfied {\em only~} if $W_1^\sigma$ and $W_2^\sigma$ are chosen to depend
on $Q^2$ and $\nu$ off shell.  In contrast Bodek and Ritchie\cite{br}, in
a study of Fermi smearing at high $Q^2$, assumed $W_1^\sigma$ and
$W_2^\sigma$ to be functions of $Q^2$ and $W$ off shell, which is
equivalent to choosing $Q^2$ and $\nu$ and evaluating $\nu$ at the point
$\nu_W\equiv(W^2-M^2+Q^2)/2M$.  Since off shell $\nu_W\not=\nu$ in general,
this prescription will not satisfy (\ref{crucial}) and implies $W_L^A\not=0$
at $Q^2=0$.  We therefore believe that their prescription is not useful
when seeking an off-shell extrapolation valid at low $Q^2$.  (Except for
the region $Q^2\simeq0$, the effect of the two choices differs little.)

When considering inelastic nucleons on shell it is convenentional to
express $W_1^\sigma(Q^2,\nu)$ in terms of $W_2^\sigma(Q^2,\nu)$ and a
third function $R_\sigma(Q^2,\nu)$, which is proportional to the ratio
$W_L/W_T$:

        \begin{equation}W_1^\sigma(Q^2,\nu)=
        {1+\nu^2/Q^2\over 1+R_\sigma(Q^2,\nu)}W_2^\sigma(Q^2,\nu)\ .
        \label{rdef}\end{equation}

\noindent  With the above assumptions for $W_1^\sigma$ and $W_2^\sigma$,
inserting (\ref{w1a})--(\ref{rdef}) into (\ref{wldef}) and (\ref{wtdef})
gives

        \begin{equation}W_L^A(\omega,{\bf q})=2\ \sum_\sigma\int
        {d^3p\over(2\pi)^3}{n_\sigma({\bf p})\over E_{\bf p}/M}
        \Biggl[{{\bf q}^2\over Q^2}
        \biggl(1+{p_z\over M}{Q^2\over|{\bf q}|\nu}\biggr)^2
        \biggl({\nu\over\omega}\biggr)^2-{1+{\nu}^2/Q^2\over
	1+R_\sigma(Q^2,\nu)}\Biggr]
        W_2^\sigma(Q^2,\nu)\ ,\label{wlar}\end{equation}

        \begin{equation}W^A_T(\omega,{\bf q})=2\times 2\ \sum_\sigma\int
        {d^3p\over(2\pi)^3}{n_\sigma({\bf p})\over E_{\bf p}/M}
        \Biggl[{1+{\nu}^2/Q^2\over 1+R_\sigma(Q^2,\nu)}
        +\biggl({p_x\over M}\biggr)^2\Biggr]
        W_2^\sigma(Q^2,\nu)\ .\label{wtar}
        \end{equation}

\noindent From (\ref{rdef}) we see that in order to satisfy (\ref{crucial})
we must have $R_\sigma(Q^2,\nu)=0$ at $Q^2=0$.  In Section 5 we provide a
parameterized form for $R_\sigma(Q^2,\nu)$ which has this property.

\section{Spectator models of nuclear binding}
\setcounter{equation}{0}
\seceqd

In order to evaluate (\ref{wlar}) and (\ref{wtar}) for bound nucleons
we must specify a particular model which determines the off-shell
kinematics.  In conventional terminology, a ``spectator'' model
specifies the energy $p_0$ of a bound nucleon in terms of its
three-momentum ${\bf p}$, which is related to the excitation energy
of the recoiling residual $(A\!-\!1)$-body target nucleus (spectator).
This in turn determines the off-shell value of the variable
$\nu=p\cdot q/M$.  As in the PWIA of (\ref{smear}), a spectator
model is based on the assumption that nuclear interactions enter
only the initial state; the final excited nucleon state is free.
In this section we discuss three such models for inelastic nucleon
response.

Bodek and Ritchie\cite{br} have investigated the effects of Fermi motion
and nuclear binding beginning with (\ref{smear}).  In that work they
assumed a nucleon energy of the form

        \begin{equation}p_0=M_A-\sqrt{{\bf p}^2+M_{A-1}^2}\ ,
	\label{p0bod}\end{equation}

\noindent where $M_A=A(M-E/A)$ is the mass of the $A$-body target nucleus
and $M_{A-1}=M_A-M$ is the mass of the recoiling spectator nucleus after
nucleon knockout.  This form is a direct generalization of the expression
for dissociation of the deuteron, as given by Atwood and West\cite{awest},
and corresponds roughly to the ``separation'' energy, i.e., the energy
required to remove a particle from the least bound state in the target.
Compared to the free nucleon energy $p_0=E_{\bf p}$, which is bounded
by $M\le E_{\bf p}\le M+34.5{\rm\ MeV}$ on the range $0\le|{\bf p}|\le
p_F$, (\ref{p0bod}) has relatively weak dependence on ${\bf p}$, being
bounded by $M\le p_0\le M-0.6{\rm\ MeV}$ over the same range.\footnote
{The numbers quoted in this section are based on numerical parameters
for ${}^{56}{\rm Fe}$ given in Section 6.}  Expression (\ref{p0bod})
has two peculiar features when applied to a many-body system: 1) $p_0$
decreases with increasing ${\bf p}$, and 2) $p_0$ includes essentially
no binding effects, since $p_0=M$ at ${\bf p}=0$ and for ${\bf p}\not=0$
the binding energy $E/A$ enters only negligibly.  Furthermore, because
(\ref{p0bod}) corresponds to the separation energy it does not account
for the possibility that the ejected nucleon was initially deeply bound.

A simple modification of (\ref{p0bod}) is to account for the variation
of $p_0$ with ${\bf p}$ using a self-consistent potential model.  We
employ a relativistic mean field model based on quantum hadrodynamics
(QHD)~\cite{sw}, a relativistic quantum field theory of hadronic matter
with Lorentz vector and scalar meson interactions.  In this model the
energy of a bound nucleon is given by

        \begin{equation}p_0=V_0+\sqrt{{\bf p}^2+{M^*}^2}\ ,
	\label{p0qhd}\end{equation}

\noindent where $M^*\equiv M+S$ is the effective nucleon mass, $S<0$
represents an attractive scalar field and $V_0>0$ represents a repulsive,
time-like vector field.  The binding effects are much greater than in
(\ref{p0bod}).  The energy $p_0$ in (\ref{p0qhd}) is bounded by
$M-76{\rm\ MeV}\le p_0\le M-24{\rm\ MeV}$ over the range
$0\le |{\bf p}|\le p_F$, i.e., nucleons in the Fermi sea are bound
by approximately $50{\rm\ MeV}$.  Furthermore, $p_0$ has the expected
behavior, i.e., {\it increases} with increasing {\bf p}, and correctly
accounts for the average binding energy for nucleons bound deeply in
the Fermi sea.  In this sense (\ref{p0qhd}) corresponds to the ``removal''
energy from occupied orbitals in the target.

Simply using (\ref{p0qhd}) to evaluate $\nu$ in (\ref{wlar}) and
({\ref{wtar}) is not consistent, however, because (\ref{wuv}) was derived
for a nucleon of mass $M$.  In the mean field QHD theory, a bound nucleon
acquires an effective mass $M^*$, on which its Lorentz transformation
properties are based.  For a nucleon of mass $M^*$ (\ref{wuv}) must be
replaced by

        \begin{equation}W_{\mu\nu}^\sigma(p,q)=W_1^\sigma
        \Biggl[-g_{\mu\nu}+{q_\mu q_\nu\over q^2}\Biggr]
        +W_2^\sigma\Biggl[{p^*_\mu\over M^*}-{p^*\!\cdot q\over M^*}
        {q_\mu\over q^2}\Biggr]\Biggl[{p^*_\nu\over M^*}
        -{p^*\!\cdot q\over M^*}{q_\nu\over q^2}\Biggr]\ ,
        \label{wuvs}\end{equation}

\noindent where $p^*\equiv(E_{\bf p}^*,{\bf p})$ and $E_{\bf p}^*\equiv
\sqrt{{\bf p^2}+{M^*}^2}$.  (Notice that the vector field $V_0$ does
not enter the definition of $p^*$, just as it does not enter a
spinor $u_s({\bf p})$ describing a Dirac plane-wave.)  It is then
natural to introduce the variable $\nu^*\equiv p^*\!\cdot q/M^*$.
Repeating the arguments of Section 3, we assume that off-shell
$W_1^\sigma$ and $W_2^\sigma$ are given by the on-shell functions
evaluated at $Q^2$ and $\nu^*$.  This leads to the following expressions,
which replace (\ref{wlar}) and (\ref{wtar})

        \begin{equation}W_L^A(\omega,{\bf q})=2\ \sum_\sigma\int
        {d^3p\over(2\pi)^3}{n_\sigma({\bf p})\over E^*_{\bf p}/M^*}
        \Biggl[{{\bf q}^2\over Q^2}
        \biggl(1+{p_z\over M^*}{Q^2\over|{\bf q}|\nu^*}\biggr)^2
        \biggl({\nu^*\over\omega}\biggr)^2-{1+{\nu^*}^2/Q^2\over
	1+R_\sigma(Q^2,\nu^*)}\Biggr]
        W_2^\sigma(Q^2,\nu^*)\ ,\label{wlas}\end{equation}

        \begin{equation}W^A_T(\omega,{\bf q})=2\times 2\ \sum_\sigma\int
        {d^3p\over(2\pi)^3}{n_\sigma({\bf p})\over E^*_{\bf p}/M^*}
        \Biggl[{1+{\nu^*}^2/Q^2\over 1+R_\sigma(Q^2,\nu^*)}
        +\biggl({p_x\over M^*}\biggr)^2\Biggr]
        W_2^\sigma(Q^2,\nu^*)\ .\label{wtas}
        \end{equation}

\noindent These are the primary formulae used in our calculations of the
inelastic nucleon background.  These expressions are also used to compute
the elastic nucleon response, i.e., quasielastic peak, for a bound Fermi
gas.  Numerical results, along with essential differences in interpretation,
are given in Section 6.

\section{Nucleon structure functions}
\setcounter{equation}{0}
\seceqe

In this section we describe the forms of the nucleon structure functions
$W_{\mu\nu}^\sigma$ used in our calculations of the inelastic nucleon
background.  We rely on two parametric fits \cite{bodek,whitlow} which
characterize $(e,e^\prime)$ data obtained over a large kinematic range
at SLAC, but which unfortunately do not cover all of the lower energy
and momentum transfers available at CEBAF.  We therefore must extrapolate
these fits to this lower range, which introduces some uncertainty into
the resulting estimates, as we shall see by comparing the results for
different fits.

The two experimental structure functions are often presented in terms of
$W_2^\sigma$ and $R_\sigma$, as in (\ref{rdef}), and given in terms of
the invariant variables $Q^2$ and $\nu$, or $Q^2$ and $x\equiv Q^2/2M\nu$;
the latter choice is of interest for describing scaling behavior at high
$x$.  Data for the proton functions comes directly from ${}^1{\rm H}
(e,e^\prime)X$, while the neutron functions must be extracted from
${}^2{\rm H}(e,e^\prime)X$ data using theoretical assumptions about
the momentum distribution and binding effects in the deuteron.

Bodek et.~al.~\cite{bodek} fit data for $W_2^\sigma$ in the kinematic
range $1.0\le Q^2\le 20{\rm\ GeV}^2$ and $0.1\le x\le 0.77$.  They assume
constant $R_p=R_n=0.18$, which allows (\ref{dcs2}) to be expressed entirely
in terms of $W_2^\sigma$.  Following a treatment developed by
West\cite{west,awest}, they use deuteron wavefunctions to account for
Fermi smearing and binding effects in extracting the neutron structure
function.  The parameterization for both proton and neutron structure
functions is then given in the form

        \begin{equation}W_2^\sigma(Q^2,\nu)=
        B(Q^2,W)\ {\tilde W}_2^\sigma(Q^2,\nu)\ ,
        \label{w2bodek}\end{equation}

\noindent where $B(Q^2,W)$ is a modulating function for low energy
excitations including threshold and resonant behavior, as a function
of final invariant mass $W=(M^2+2M\nu-Q^2)^{1/2}$.  We eliminate the
resonant terms from this function since they do not appreciably modify
the estimates, and since the value of $R$ is expected to vary from
resonance to resonance, contrary to the assumption of constant $R_p=R_n$.
Without resonances, $B(Q^2,W)$ vanishes below the $\pi$-production
threshold $W\simeq 1.07{\rm\ GeV}$, increases with increasing $W$ to
a second threshold near $W\simeq 1.74{\rm\ GeV}$, and reaches unity for
$W\rsim 2M$; the functional form of $B(Q^2,W)$ is given in Appendix A.
The function ${\tilde W}_2^\sigma$ in (\ref{w2bodek}) is a smoothly
varying function of $Q^2$ and $\nu$, and is also given in Appendix A.
Its functional form was chosen to illustrate the behavior of
$W_2^\sigma(Q^2,\nu)$ in a certain scaling variable (see
Ref.~\cite{bodek}).

Whitlow\cite{whitlow} fits $W_2^\sigma$ in the kinematic range
$0.6\le Q^2\le 30{\rm\ GeV}^2$ and $0.06\le x\le 0.9$, using all
SLAC data available at the time of publication.  He first extracts
$R_\sigma(Q^2,x)$ using only data for which $W>2{\rm\ GeV}$ to exclude
the resonance region.  He finds $R_p=R_d$, and therefore $R_p=R_n$, to
within experimental errors ($\sim 5\%$), in good agreement with theoretical
predictions based on QCD.\cite{altarelli}  However, noting that QCD
predictions\cite{altarelli,georgi} systematically underestimate the
data, he provides a parameterization of the experimentally determined
$R_\sigma$ in the form

        \begin{equation}R_\sigma(Q^2,x)={1\over3}
        \biggl[R_a(Q^2,x)+R_b(Q^2,x)+R_c(Q^2,x)\biggr],
        \quad{\rm for}\ Q^2\ge 0.3\ ,
        \label{rwhitlow}\end{equation}

\noindent where $R_a$, $R_b$ and $R_c$ are given in Appendix B.

For fixed $x$ and $Q^2\rsim 1{\rm\ GeV^2}$, the experimental data
indicates that $R_\sigma\simeq 0.2$ and decreases gradually as $Q^2$
increases.  For $Q^2\lsim 1{\rm\ GeV^2}$, $R_\sigma$ decreases rapidly
and tends to zero, as required to ensure $W_L=0$ at $Q^2=0$.  In contrast,
if (\ref{rwhitlow}) is evaluated for $Q^2<0.3{\rm\ GeV^2}$ it will
diverge as $Q^2\rightarrow0$.  To make reasonable estimates, therefore,
we extrapolate $R_\sigma(Q^2,x)$ down to $Q^2=0$ by assuming a simple
analytic form which behaves properly at all $Q^{2}$.  A theoretically
motivated form\cite{unki} is

        \begin{equation}R_\sigma(Q^2,x)={a_R(x)\ Q^2\over b_R(x)+Q^4},
        \quad{\rm for}\ Q^2 < 0.3\ ,\label{twist}\end{equation}

\noindent where $a_R(x)$ and $b_R(x)$ are chosen such that (\ref{twist})
smoothly matches (\ref{rwhitlow}) at $Q^2=0.3{\rm\ GeV^2}$.  We have
verified that the results in Section 6 are insensitive to the detailed
form of this continuation, provided it matches (\ref{rwhitlow}) and tends
smoothly to zero at $Q^2=0$.

The proton structure function $W_2^p$ is parameterized in the form

        \begin{equation}W_2^p(Q^2,x)={F_2^{\rm thr}(x)\over Q^2/2Mx}
	\Biggl[1+\lambda_1(x)\ {\rm ln}\biggl[{Q^2\over A(x)}\biggr]
	+\lambda_2(x)\ {\rm ln}^2\biggl[{Q^2\over A(x)}\biggr]
	\Biggr]\ ,\label{w2whitlow}\end{equation}

\noindent where explicit forms for the functions appearing here are
given in Appendix C.  The dominant behavior of (\ref{w2whitlow}) is
determined by the function $F_2^{\rm thr}(x)$, which is a power series
in $(1-x)$ and therefore tends to zero at the elastic nucleon threshold
at $x=1$.  The neutron structure function is parameterized in terms of
the ratio

        \begin{equation}{W_2^n\over W_2^n}\simeq
        \left({W_2^n\over W_2^n}\right)_{\rm S}
        =P_1(x)+P_2(x)\ \Bigl[{\rm\ ln}\ Q^2\Bigr]\ ,
        \label{w2nw2p}\end{equation}

\noindent where the subscript S refers to Fermi smearing effects in
the deuteron, which Whitlow points out are negligible for $x<0.5$.
In this region the data looks almost linear in $x$, although strong
curvature is apparent for $0.5<x<0.7$.  This curvature is most likely
an artifact of Fermi smearing in the deuteron.  Whitlow gives two
different parameterizations of the functions $P_1(x)$ and $P_2(x)$,
one essentially linear in $x$ and one which shows some curvature
corresponding to this Fermi motion effect.  Bodek et.~al.~\cite{bodek}
have extracted the non-smeared ratio $W_2^n/W_2^n$ and find linear
$x$-dependence over their whole range of data.  For the purpose of
our estimates, therefore, we choose the more linear of Whitlow's two
parameterizations even though it was fit only for $x<0.5$.  Explicit
forms for the functions $P_{1}(x)$ and $P_{2}(x)$ which are used in
(\ref{w2nw2p}) are given in Appendix C.

Before proceeding to our estimate of the inelastic nucleon background,
it is useful to point out some general features of the fits to
$W_2^\sigma$, since these largely determine the behavior of the
nuclear response functions (\ref{wlas}) and (\ref{wtas}).  Although
data for $W_2^\sigma$ are shown in the original references, they are
typically shown at higher energies and in terms of invariant variables,
e.g., $Q^2$ and $x$.  Figures 1--3 show $W_2^p$ and $W_2^n$ in laboratory
variables $\omega$ and $|{\bf q}$|, for free nucleons at rest and for
three-momentum transfers $|{\bf q}|=1,\ 2\ {\rm and}\ 4{\rm\ GeV/c}$.
The energy range shown is $E_{\bf q}-M\le\omega\le |{\bf q}|$, i.e.,
from the position of the elastic nucleon peak to the absolute upper
limit attainable by electron scattering.

We first focus on protons.  At these relatively low momenta, the fit
by Bodek et.~al.~(B) is significantly smaller than that by Whitlow
($\Lambda_{12}$), with nearly a factor of two difference for
$|{\bf q}|=1{\rm\ GeV/c}$.  This difference decreases with increasing
$|{\bf q}|$, and by $|{\bf q}|=4{\rm\ GeV/c}$ the two fits are nearly
indistinguishable.  At $|{\bf q}|=2{\rm\ GeV/c}$, B shows the second
(higher) threshold behavior of (\ref{btil}), while at $|{\bf q}|=1
{\rm\ GeV/c}$ the values of the final invariant mass $W$ are such that
all $\omega$ lie entirely below the second threshold.  The behavior of
B and $\Lambda_{12}$ at the endpoints differs qualitatively.  At the
lower limit, i.e., the elastic threshold, $\Lambda_{12}$ vanishes at
$\omega=\sqrt{{\bf q}^2+M^2}-M$ (see discussion following
(\ref{w2whitlow})), while B vanishes at the first inelastic threshold
$\omega=\sqrt{{\bf q}^2+W^2}-M$ with $W\simeq1.07{\rm\ GeV/c}$.  At
the upper limit B vanishes precisely at $\omega=|{\bf q}|$, while the
logarithmic behavior of $\Lambda_{12}$ in $Q^2$ must be cut off to
allow only positive values, causing $W_2^p$ to vanish slightly prior
to $\omega=|{\bf q}|$.  Generally speaking, the fits to $W_2^n$ follow
similar trends, although with different numerical values.  For both
fits the neutron-to-proton ratio $W_2^n/W_2^p<1$.  Compared to the B
fit, this ratio in the $\Lambda_{12}$ fit, i.e., that given by
(\ref{w2nw2p}) is $\sim40\%$ smaller at the lower limit and $\sim10\%$
larger at the upper limit for $|{\bf q}|=1{\rm\ GeV/c}$, and is only
$\sim25\%$ smaller at the lower limit and $\sim5\%$ larger at the upper
limit for $|{\bf q}|=2{\rm\ GeV/c}$.  In Fig.~1, the difference in shape
between $W_2^n$ and $W_2^p$ in the $\Lambda_{12}$ fit is reflective of
the fact that at lower $|{\bf q}|$ the slope of the ratio $W_2^n/W_2^p$
is greater than that in the B fit.

\section{Numerical results}
\setcounter{equation}{0}
\seceqf

In this section we present numerical results for the inelastic nucleon
background obtained by using (\ref{wlas}) and (\ref{wtas}) with the
fits to $W_{2}^{\sigma}$ and $R_{\sigma}$ given in Section 5.  We
assume a simple Fermi gas momentum distribution $n_{\sigma}({\bf p})
=\theta(p_F-|{\bf p}|)$ with $p_F=0.257{\rm\ fm}^{-1}$, corresponding
roughly to ${}^{56}{\rm Fe}$.  The free nucleon mass $M=938.92{\rm\ MeV}$.
To account for nuclear binding we use a reduced mass $M^{*}/M=0.648$,
corresponding to the mean field theory of Ref.~\cite{sw} for the
same value of $p_F$.  This value of $M^*$ represents the smallest
reduced mass which can reasonably be expected in a real nucleus, so
that comparing this result with that for free nucleons demonstrates
the range of results which can be expected in this model.  In all cases
where $W$ and $x$ must be replaced by the variables $Q^2$ and $\nu$ in
the functions $W_2^\sigma$ and $R_\sigma$, as mentioned in Section 5,
these replacements have been made in the on-shell functions {\em before}
letting $\nu\rightarrow\nu^*$.

In what follows the inelastic background will be shown compared to
the corresponding quasielastic peak at the same three-momentum transfer,
whose calculation for free nucleons is summarized in Appendix D.  To
include the effects of nuclear binding on the quasielastic peak we
follow the theoretical treatment of Ref.~\cite{kf}, which is described
at the end of Appendix D.  This leads to the standard expressions for
quasielastic scattering from a Fermi gas composed of nucleons of mass
$M^{*}$, for which binding effects enter {\em both} initial and final
nucleon states.  Note that this differs from the spectator models
described in Section 4, for which binding effects enter final states
only indirectly, since there is assumed to be no final state interaction
of the inelastically excited nucleon.  Therefore, the comparison between
the elastic and inelastic nucleon contributions computed here is given
only to provide a reasonable estimate of their relative sizes and
positions.

Figure 4 shows the longitudinal (a) and transverse (b) nuclear response
functions at three-momentum transfer $|{\bf q}|=1{\rm\ GeV/c}$.  Thin
curves are for free nucleons with mass $M$ and thick curves are for bound
nucleons with mass $M^*$.  As seen in Fig.~1, at this three-momentum
the fit by Bodek et.~al.~(B) implies a background roughly half as large
as that implied by Whitlow's fit ($\Lambda_{12}$).  (Note that the energy
range extends slightly lower than that in Fig.~1 to accommodate Fermi
broadening.)  As in Fig.~1, B vanishes at exactly $\omega=|{\bf q}|$,
while $\Lambda_{12}$ vanishes at slightly lower $\omega$ due to the
logarithmic behavior of (\ref{w2whitlow}).  Note that although the
inelastic nucleon results derived from B and $\Lambda_{12}$ differ,
they are of the same magnitude as the quasielastic peak and displaced
sufficiently that respective peaks appear at distinct energies.
It therefore seems possible to effect a separation of elastic and
inelastic nucleon contributions in this momentum range using these
data fits, albeit with substantial errors reflecting the difference
between the fits.

The reduced mass $M^{*}$ affects the quasielastic peak much more than
the inelastic background.  To understand this result we first note that
Fermi motion is treated identically in the two calculations, and
that in (\ref{wuvs}) the effective mass always enters the coefficients
of $W_{1}^{\sigma}$ and $W_2^{\sigma}$ through the ratio $p_{0}^{*}/M^{*}
=E_{\bf p}^{*}/M^{*}\sim1$.  Thus differences in sensitivity to the value
of $M^{*}$ originate in the nucleon structure functions $W_{1}^{\sigma}$
and $W_2^{\sigma}$.  We have chosen to evaluate the inelastic nucleon
structure functions in terms of $Q^{2}$ and $\nu^{*}\sim\omega$, hence
the inelastic nucleon background is not strongly sensitive to the value
of $M^{*}$.  It is in this sense that our background calculation
includes binding effects only indirectly in the final state.  In
contrast, in order to match standard treatments the elastic nucleon
structure functions (\ref{w1el}) and (\ref{w2el}) have explicit factors
of $\tau^{*}\equiv Q^2/4{M^*}^2$, which scales quadratically with the
reduced mass $M^*$.  Hence the size and location of the quasielastic peak
is more sensitive to the value of $M^{*}$ than is the inelastic nucleon
background.   This is clearly an artifact of the particular model we
have used for the quasielastic peak, and would change with any
modification of the theory for either elastic or inelastic nucleons.
Such changes are not of interest here, since our main goal is to provide
a reasonable estimate the inelastic background and determine how reliably
these parameterizations of the SLAC data can be extended to the CEBAF
kinematic range.

Figure 5 shows the longitudinal (a) and transverse (b) nuclear response
functions at three-momentum transfer $|{\bf q}|=2{\rm\ GeV/c}$.  The
curve labeling is the same as in Fig.~4.  The threshold behavior of B,
which can be seen in Fig.~2, is not visible in Fig.~5 because of Fermi
smearing.  The most notable difference with Fig.~4 is that the inelastic
background is {\em much} larger compared to the quasielastic peak.
The scales of the plots show that this is due mainly to the rapid
decrease of the dipole form factor (\ref{gep}), and due to a lesser
extent to a slight increase in the inelastic background.  Thus
separating elastic and inelastic nucleon contributions from experimental
data will be much more challenging for $|{\bf q}|\rsim2{\rm\ GeV/c}$
-- an effect which has long been recognized but not quantitively
investigated.  However, this difficulty is partly compensated by the
fact that, as noted in Fig.~2, the two inelastic fits B and $\Lambda_{12}$
are in much closer agreement, and therefore the inelastic background
can be evaluated with more confidence.  Results for $|{\bf q}|=4
{\rm\ GeV/c}$ are not shown because by this point the dipole form
factor has totally suppressed the elastic peaks relative to the
inelastic background.

 \section{Summary and conclusions}

The study of nuclear structure by $(e,e^\prime)$ reactions at a few
GeV, the CEBAF kinematic range, requires the removal of the background
resulting from inelastic excitation of single nucleons.  In this paper
we provide an estimate of this background based on two different
parameterizations of inelastic nucleon structure functions measured
at SLAC.  We then compare the sizes and positions of the inelastic
nucleon background to those of the quasielastic peak, which represents
the dominant contribution of elastic nucleons to inelastic nuclear
excitation.  We assess how confidently the available SLAC fits can be
applied in the CEBAF kinematic range from the difference between the
results for each fit over that kinematic range.

We make a number of assumptions which allow us to relate inelastic
nucleon excitations in the nuclear target to those of free nucleons;
these all involve allowing only ``minimal'' effects of the nucleus
on internal nucleon excitations.  First we use the PWIA in (\ref{smear}),
which introduces Fermi motion but does not include final state
interations for the excited nucleon.  Second we assume that the
nucleon response tensor $W_{uv}^\sigma$ for a bound nucleon, given
in (\ref{wuvs}) is of the same tensor form as that for a free nucleon,
given in (\ref{wuv}), but modified for a nucleon with effective mass
$M^*$.  Third we assume that the nucleon structure functions
$W_1^\sigma$ and $W_2^\sigma$ for bound nucleons are equal to the
free-nucleon structure functions evaluated at the same value of
$Q^2$ and at $\nu=\nu^*$ (see discussion leading to (\ref{wlas})
and (\ref{wtas})).  Thus binding effects enter the enter final
states in (\ref{wuvs}) only indirectly, which is consistent with
ignoring final state interactions in (\ref{smear}).  This rule for
off-shell extrapolation has good behavior at $Q^2=0$, as seen in
(\ref{crucial}).

Our calculation of the inelastic response functions uses parameters
based on the structure of ${}^{56}{\rm Fe}$, but with a sharp Fermi
distribution.  There is little sensitivity of the inelastic nucleon
background to the choice of $M^*$, which enters only minimally.  In
contrast, our model for the quasielastic peak, which does have final
state interactions, is more sensitive to $M^*$, but this does not
significantly affect our comparison of the relative size and position
of the two contributions.  We find that for $|{\bf q}|\lsim 1
{\rm\ GeV/c}$ the elastic and inelastic nucleon contributions are
comparable in size, and can be separated, but in the inelastic nucleon
response there is an uncertainty of roughly a factor of two resulting
from disagreement between the SLAC fits in this kinematic range.  For
$|{\bf q}|\rsim 2{\rm\ GeV/c}$ improved agreement between the two
extrapolations raises the certainty of the background calculation,
but the momentum dependence of the Sachs form factors quenches the
elastic nucleon contribution with increasing $|{\bf q}|$.

Our estimates of the inelastic nucleon background show that with the
presently available SLAC fits it is feasible to extract nuclear
information from nuclear $(e,e^\prime)$ data for $|{\bf q}|\lsim 1
{\rm\ GeV/c}$.  However, the present accuracy does not make this
sufficiently useful, since many interesting effects (e.g., of
correlations) make rather small contributions to the response functions.
What is needed is more complete data and analysis on nucleon structure
functions at lower $Q^2$, with full separation of $W_1^\sigma$ and
$W_2^\sigma$, i.e., $R_\sigma$.  Clearly resonances are important
in this kinematic region, and should be included in the inelastic
nucleon background.  In contrast, for $|{\bf q}|\rsim 2{\rm\ GeV/c}$
separating the inelastic nucleon background can be done with more
confidence, but the momentum dependence of the Sachs form factors
makes identification of the elastic nucleon response difficult.
It seems that this situation can not be improved by making more
precise measurements of the inelastic nucleon response functions,
since the existing fits are already in good agreement in this
kinematic range.  For some purposes, the use of $(e,e^\prime p)$
and $(e,e^\prime n)$ experiments would reduce the problem of the
background, but these reactions are usually less complete kinematically
than $(e,e^\prime)$, and are more sensitive to final state interactions.
These features limit their usefulness in sum rule studies.

\vskip 0.3 true in
\noindent{\Large\bf Acknowledgements}
\vskip 0.2 true in

This research was supported in part by the U.S.~Department of Energy under
Grant No.~DE-FG02-88ER40425 with the University of Rochester.  The authors
would like to thank Un-Ki Yang, Arie Bodek and Steve Rock for help and
useful information regarding the SLAC data and analysis, and the High
Energy Physics Group for use of their VAX computer.

\vskip 0.3 true in
\noindent{\Large\bf Appendix A:\ \ \ Fit to $W_2^\sigma(Q^2,\nu)$
by Bodek et.~al.}
\vskip 0.2 true in
\setcounter{equation}{0}
\seceqaa

In Ref.~\cite{bodek} the proton and neutron structure functions are
parameterized in the form

        \begin{equation}{\tilde W}_2^\sigma(Q^2,\nu)=
        \Biggl[{Q^2\over2M\nu^2}
        {2M\nu+c_1\over Q^2+c_2}\Biggr]
        \sum_{n=3}^7c_{n\sigma}
        \ \Biggl[1-{Q^2+c_2\over2M\nu+c_1}
        \Biggr]^n\ ,\label{global}\end{equation}

\noindent where $c_1=+1.6421$ and $c_2=+0.3764$.  The remaining
parameters $c_{n\sigma}$ are listed in Table 1.

\vskip 0.25 true in
\centerline{
\begin{tabular}{|c|c|c|}
	\hline
	 & $\sigma=p$ & $\sigma=n$  \\
	\hline
	$c_{3\sigma}$ & $+0.2562$ & $+0.0640$  \\
	\hline
	$c_{4\sigma}$ & $+2.1785$ & $+0.2254$  \\
	\hline
	$c_{5\sigma}$ & $+0.8978$ & $+4.1062$  \\
	\hline
	$c_{6\sigma}$ & $-6.7162$ & $-7.0786$  \\
	\hline
	$c_{7\sigma}$ & $+3.7557$ & $+3.0549$  \\
	\hline
\end{tabular}}
\vskip 0.25 true in
\noindent {\bf Table 1.} Numerical parameters for ${\tilde W}_2^\sigma$,
taken from Bodek {\em et.~al.~}\cite{bodek}.

\noindent The modulating function $B(Q^2,W)$ is given by

	\begin{equation}B(Q^2,W)={\tilde B}(W)
	\Bigl[1+\Bigl(1-{\tilde B}(W)\Bigr)
	\Bigl(b_6+b_7(x-b_8)^2\Bigr)\Bigr]\ ,
	\label{bmod}\end{equation}

\noindent where $x=Q^2/(W^2-M^2+Q^2)$ and ${\tilde B}(W)$ is defined

        \begin{equation}{\tilde B}(W)=
	\theta(W-b_1)\ b_2\biggl[1-{\rm e}^{-b_3(W-b_1)}\biggr]+
	\theta(W-b_4)\ (1-b_2)\biggl[1-{\rm e}^{-b_5(W^2-b_4^2)}
	\biggr]\ .\label{btil}\end{equation}

\noindent The coefficients $b_i$ are listed in Table 2.

\vskip 0.25 true in
\centerline{
\begin{tabular}{|c|c|}
	\hline
	$b_1=+1.0741$ & $b_2=-0.7553$ \\
	\hline
	$b_3=+3.3506$ & $b_4=+1.7447$ \\
	\hline
	$b_5=+3.5102$ & $b_6=-0.5999$ \\
	\hline
	$b_7=+4.7616$ & $b_8=+0.4117$ \\
	\hline
\end{tabular}}
\vskip 0.25 true in
\noindent {\bf Table 2.} Numerical parameters for $B(Q^2,W)$, supplied
by Bodek\cite{bodek2}.

\vskip 0.3 true in
\noindent{\Large\bf Appendix B:\ \ \ Fit to $R(Q^2,x)$ by Whitlow}
\vskip 0.2 true in
\setcounter{equation}{0}
\seceqab

The functions appearing in (\ref{rwhitlow}) are given by

        \begin{equation}R_a(Q^2,x)={a_1\over\ln(Q^2/0.04)}
        \Theta(Q^2,x)+{a_2\over\Bigl(Q^8+a_3^4\Bigr)^{1/4}}\ ,
        \label{ra}\end{equation}

        \begin{equation}R_b(Q^2,x)={b_1\over\ln(Q^2/0.04)}
        \Theta(Q^2,x)+{b_2\over Q^2}+{b_3\over\Bigl(Q^4+0.3^2
        \Bigr)}\ ,\label{rb}\end{equation}

        \begin{equation}R_c(Q^2,x)={c_1\over\ln(Q^2/0.04)}
        \Theta(Q^2,x)+{c_2\over\sqrt{\Bigl(Q^2-5(1-x)^5\Bigr)^2
        +c_3^2}}\ ,\label{rc}\end{equation}

\noindent where

        \begin{equation}\Theta(Q^2,x)=1+12\biggl({Q^2\over Q^2+1}
        \biggr)\biggl({0.125^2\over x^2+0.125^2}\biggr)\ ,
        \label{theta}\end{equation}

\noindent and the coefficients appearing in (\ref{ra})--(\ref{rc}) are
listed in Table 3.

\vskip 0.25 true in
\centerline{
\begin{tabular}{|c|c|c|}
	\hline
	$a_1=+0.0672$ & $a_2=+0.4671$ & $a_3=+1.8979$  \\
	\hline
	$b_1=+0.0635$ & $b_2=+0.5747$ & $b_3=-0.3534$  \\
	\hline
	$c_1=+0.0599$ & $c_2=+0.5088$ & $c_3=+2.1081$  \\
	\hline
\end{tabular}
}\vskip 0.25 true in
\noindent {\bf Table 3.} Numerical parameters for $R(Q^2,x)$, taken
from Whitlow\cite{whitlow}.

\vskip 0.3 true in
\noindent{\Large\bf Appendix C:\ \ \ Fit to $W_2^\sigma(Q^2,x)$
by Whitlow}
\vskip 0.2 true in
\setcounter{equation}{0}
\seceqac

The functions appearing in (\ref{w2whitlow}) are given by

        \begin{equation}F_2^{\rm thr}(Q^2,x)=\sum_{i=1}^5
	d_i\ (1-x)^{i+2}\ ,\label{f2thr}\end{equation}

	\begin{equation}\lambda_1(x)=\sum_{i=0}^3
	d_{i+9}\ x^i\ ,\label{lambda1}\end{equation}

	\begin{equation}\lambda_2(x)=
	\left\{ \begin{array}{ll}
	\sum_{i=0}^2 d_{i+6}\ x^i & {\rm if\ }Q^2<A(x)\ , \\
	\qquad 0 & {\rm otherwise\ ,} \end{array}\right.
	\label{lambda2}\end{equation}

	\begin{equation}A(x)=1.22\ {\rm e}^{3.2x}\ ,
	\label{ax}\end{equation}

\noindent where the coefficients $d_i$ are listed in Table 4.

\vskip 0.25 true in
\centerline{
\begin{tabular}{|c|c|c|}
	\hline
	$d_1=+1.417$ & $d_2=-0.108$ & $d_3=+1.486$  \\
	\hline
	$d_4=-5.979$ & $d_5=+3.524$ & $d_6=-0.011$  \\
	\hline
	$d_7=-0.619$ & $d_8=+1.385$ & $d_9=+0.270$  \\
	\hline
	$d_{10}=-2.179$ & $d_{11}=+4.722$ & $d_{12}=-4.363$  \\
	\hline
\end{tabular}}
\vskip 0.25 true in
\noindent {\bf Table 4.} Parameters for Whitlow's $\Lambda_{12}$ fit
to $W_2^p(Q^2,x)$, taken from Ref.~\cite{whitlow}.

\noindent The functions $P_1(x)$ and $P_2(x)$ appearing in (\ref{w2nw2p})
are given by

	\begin{equation}P_1(x) \simeq 0.9498-0.9706\ x
	+0.3102\ x^2\ ,\label{p1x}\end{equation}

	\begin{equation}P_2(x) \simeq -0.0146\ ,
	\label{p2x}\end{equation}

\noindent where the numerical values are taken from Ref.~\cite{whitlow}.

\vskip 0.3 true in
\noindent{\Large\bf Appendix D:\ \ \ Elastic nucleon structure functions}
\vskip 0.2 true in
\setcounter{equation}{0}
\seceqad

The elastic structure functions for free nucleons may be written

        \begin{equation}W_1^\sigma(Q^2,\nu)=\tau\
        G_{M\sigma}^2(Q^2)\ \delta\biggl(\nu-{Q^2\over 2M}\biggr)\ ,
        \label{w1el}\end{equation}

        \begin{equation}W_2^\sigma(Q^2,\nu)=
        {G_{E\sigma}^2(Q^2)+\tau G_{M\sigma}^2(Q^2)
        \over 1+\tau}\ \delta\biggl(\nu-{Q^2\over 2M}\biggr)\ ,
        \label{w2el}\end{equation}

\noindent where $G_{E\sigma}(Q^2)$ and $G_{M\sigma}(Q^2)$ are the
Sachs electric and magnetic form factors, and $\tau\equiv Q^2/4M^2$.
Inserting (\ref{w1el}) and (\ref{w2el}) into (\ref{wla}) and
(\ref{wta}) leads to

        \begin{equation}W_L^A(\omega,{\bf q})={Q^2\over{\bf q}^2}
        \times 2\sum_\sigma\int{d^3p\over(2\pi)^3}\
        f_\sigma({\bf p})\ {L_{00}^\sigma({\bf p},{\bf q};Q^2)
        \over 4E_{\bf p}E_{\bf p+q}}\delta\bigl(\omega+E_{\bf p}
        -E_{\bf p+q}\bigr)\ ,\label{wlqe}\end{equation}

        \begin{equation}W_T^A(\omega,{\bf q})=2\times 2\sum_\sigma
        \int{d^3p\over(2\pi)^3}\ f_\sigma({\bf p})\
        {L_{11}^\sigma({\bf p},{\bf q};Q^2)\over 4E_{\bf p}E_{\bf p+q}}
        \delta\bigl(\omega+E_{\bf p}-E_{\bf p+q}\bigr)\ ,
        \label{wtqe}\end{equation}

\noindent where

        \begin{equation}L_{00}^\sigma({\bf p},{\bf q};Q^2)=
        {G_{E\sigma}^2(Q^2)\over 1+\tau}(E_{\bf p}+E_{\bf p+q})^2
        +{G_{M\sigma}^2(Q^2)\over 1+\tau}\Bigl[\tau(E_{\bf p}
        +E_{\bf p+q})^2-(1+\tau){\bf q}^2\Bigr]\ ,
        \label{L00}\end{equation}

        \begin{equation}L_{11}^\sigma({\bf p},{\bf q};Q^2)=
        {G_{E\sigma}^2(Q^2)\over 1+\tau}(2\ p_x)^2
        +{G_{M\sigma}^2(Q^2)\over 1+\tau}\Bigl[\tau(2\ p_x)^2+
        (1+\tau)Q^2\Bigr]\ .\label{L11}\end{equation}

\noindent The factor $f_\sigma({\bf p})\equiv n_\sigma({\bf p})
\Bigl[1-n_\sigma({\bf p+q})\Bigr]$ in (\ref{wlqe}) and (\ref{wtqe})
provides Pauli blocking for final states, and must be included when
starting from (\ref{smear}) (see footnote 1).  Expressions (\ref{wlqe})
and (\ref{wtqe}) can be evaluated numerically by rewriting the
$\delta$-function as a $\theta$-function which restricts the polar
angle $\theta\equiv{\rm cos}^{-1}\left[{\bf p \cdot q}/|{\bf p}|
|{\bf q}|\right]$.

In numerical calculations we take the Sachs form factors to be

	\begin{eqnarray}
	G_{Ep}(Q^2)&=&(1+Q^2/0.71\ {\rm GeV}^2)^{-2}\ ,\label{gep}\\
	G_{Mp}(Q^2)&=&(1+\kappa_p)G_{Ep}(Q^2)\ ,\label{gmp}\\
	G_{Mn}(Q^2)&=&\kappa_n G_{Ep}(Q^2)\ ,\label{gmn}\\
	G_{En}(Q^2)&=&0\ .\label{gen}
	\end{eqnarray}

\noindent where $\kappa_p\!=\!+1.79$ and $\kappa_n\!=\!-1.91$ are the
proton and neutron anomalous magnetic moments, respectively.

For nucleons of mass $M^{*}$ the above expressions must be modified.
This is accomplished in part by letting $E_{\bf p}\rightarrow E_{\bf p}^*$,
$E_{\bf p+q}\rightarrow E_{\bf p+q}^*$ and $\tau\rightarrow \tau^*$ in
expressions (\ref{wlqe})--(\ref{L11}).   While these changes must be
made to account for the modified kinematics of nucleons with a reduced
mass $M^*$, there is some ambiguity in the treatment of the Sachs form
factors which are fit only to free nucleon data.  In this work we adopt
Model G of Ref.~\cite{kf}, in which we assume that the Sachs electric
and magnetic form factors are unmodified in the nuclear medium.

\vfill
\eject

\noindent{\Large\bf Figure Captions}
\vskip 0.2 true in

\noindent FIG.~1.  Fits to nucleon structure functions $W_2^p$ and
$W_2^n$ in laboratory variables for three-momentum transfer
$|{\bf q}|=1{\rm\ GeV/c}$.  Fits by Whitlow ($\Lambda_{12}$) for the
proton (solid) and neutron (dashed), and fits by Bodek et.~al.~(B)
for the proton (dot-dashed) and neutron (dotted).
\hfill\break

\noindent FIG.~2.  Same as Fig.~1 except for $|{\bf q}|=2{\rm\ GeV/c}$.
\hfill\break

\noindent FIG.~3.  Same as Fig.~1 except for $|{\bf q}|=4{\rm\ GeV/c}$.
\hfill\break

\noindent FIG.~4.  Nuclear response functions (per nucleon) for
three-momentum transfer $|{\bf q}|=1{\rm\ GeV/c}$: (a) longitudinal
$W_L^A(\omega,{\bf q})/A$, and (b) transverse $W_T^A(\omega,{\bf q})/A$.
Inelastic nucleon background based on fits by Whiltow ($\Lambda_{12}$)
(solid curves) and by Bodek et.~al.~(dashed curves) are shown, along
with the quasielastic peak (dotted curves).  Thin lines are for free
nucleons with $M*/M=1$, and thick lines are for interacting nucleons
with $M*/M=0.648$.
\hfill\break

\noindent FIG.~5.  Same as Fig.~4 except for $|{\bf q}|=2{\rm\ GeV/c}$.
The inserts show enlarged views of the quasielastic peak region.
\hfill\break

\vfill
\eject

\end{document}